\documentclass[]{revtex4}
\usepackage{amsmath}
\usepackage{amssymb}
\usepackage{graphicx}

\begin{document}
\title{Effects of localized trap-states and
corrugation on charge transport in graphene nanoribbons}

\author{Oleksiy Roslyak$^{1}$,
Upali Aparajita$^{1}$, Godfrey Gumbs$^{1,2}$, and Danhong Huang$^3$}
\affiliation{$^1$Department of Physics and Astronomy,
Hunter College, City University of New York 695 Park Avenue,
New York, NY 10065, USA\\
$^2$Donostia International Physics Center (DIPC), P de
Manuel Lardizabal, 4, 20018 San Sebastian, Basque Country, Spain}
\affiliation{$^3$Air Force Research Laboratory (ARFL/RVSS),
Kirtland Air Force Base, NM 87117, USA}

\date{\today}

\begin{abstract}
We investigate the role played by electron traps on adiabatic
charge transport for graphene nanoribbons in the presence of an
acoustically induced longitudinal surface acoustic wave (SAW) potential.
Due to the weak longitudinal SAW-induced potential as well as the strong transverse confinement by a nanoribbon,
minibandsof sliding tunnel-coupled quantum dots are formed so that by varying
the chemical potential to pass through the minigaps,
quantized adiabatic charge transport may be obtained. We
analyze the way that the minigaps may be closed,
thereby destroying the likelihood of current quantization
in a nanoribbon. We present numerical calculations showing
the effects due to electron traps which lead to localized-trap
energy levels within the minigaps. Additionally, for comparison,
we present results for the minibands of a
corrugated nanoribbon in the absence of a SAW.
\end{abstract}
\maketitle

\section{Introduction}
\label{sec1saw}

A considerable amount of research work has been carried out so far
on the design and improvement of electronic devices which
are based on the use of quantized adiabatic charge
transport.\,\cite{11saw,1saw,2saw,4saw,5saw,6saw,7saw,8saw,9saw,10saw}
Moreover, under a surface-acoustic wave (SAW), the inelastic capture and tunneling escape effects on the non-adiabatic transport of photo-excited charges
in quantum wells was also investigated.\,\cite{add}
The underlying challenge is to produce a device with an accuracy
for the quantized current of one part in $10^8$ on the plateaus.
When this goal is achieved, one  application of this device
would be in metrology  for standardizing the unit of current.
\medskip

At the present time,  a SAW is launched on a piezoelectric
heterostructure, such as GaAs/AlGaAs, and GHz single/few-electron pumps
have been gaining close scrutiny due to the fact that the measured
currents lie within the nanoamp range, high enough for the measured
current to be suitable as a current standard. However, these
pumps have so far been capable of delivering electrons/holes in
each cycle of a sliding dynamic quantum dot (QD), giving rise to a quantized
current with an accuracy of one part in $10^6$ as reported in
Refs.\,[\onlinecite{2saw,4saw,5saw,6saw,7saw,8saw}].
Interestingly, in Ref.\,[\onlinecite{12saw}]
a measurement was carried out of the noise accompanying a $3$-GHz
SAW pump. It was observed in this experiment that the current
near the lowest plateau, corresponding to the transfer of one
electron per SAW cycle, is dominated by shot noise. However,
away from the plateau, the noise is attributed to electron traps
in the material. There have been some attempts to increase the
flatness of the plateaus by applying
magnetic fields.\,\cite{wright1,wright2}
\medskip

Some time ago,  a proposal was put forward by Thouless\,\cite{10saw}
which would make use of  quantized adiabatic charge
transport. This adiabatic approach involves the use of
a one-dimensional (1D) electron system
subjected to a slowly-sliding periodic potential.
Relatively simple analysis indicates that in such a 1D system
minigaps are generated in instantaneous electronic spectra as a
function of the SAW amplitude. With the use of a gate, the
chemical potential can be varied by applying a voltage to the gate.
Consequently, when the chemical potential lies within a minigap,
there will be an integral multiple of electron charge transported
across the system during a single time period.\,\cite{11saw}
In other words, by combining with the strong transverse confinement of a nanoribbon,
the weal longitudinal SAW potential has induced a series of dynamic
(sliding) tunnel-coupled QDs whose impenetrable``wall  is constructed through
destructive interference of the electronic wave functions
around a minimum of the SAW potential. In principle, such
an adiabatic-transport device could provide an important
application, like a current standard. Talyanskii,
{\em et al.\/}\,\cite{11saw} investigated the physical
mechanisms of quantized adiabatic charge transport in carbon
nanotubes for a  SAW to produce a  periodic potential required
for miniband/minigap formation.
\medskip

In the presence of a SAW, the scattering effects from impurities
embedded in a 1D electronic system are expected to play an important
role on the flatness of a current plateau. The current quantization
should be completely smeared out when the level broadening from
impurity scattering becomes comparable to the minigaps of dynamic tunnel-coupled QDs.
On the other hand, we can also simulate localized electron traps
by superposing a series of negative $\delta$-potentials onto
a SAW potential within each spatial period. Consequently,
we expect a set of localized trap states occurring within
the minigaps of dynamic QDs. This provides an escape channel
for the QD-confined electrons being carried by the SAW. This
trap mechanism is quite different from the impurity
one\,\cite{11saw} where a spatial average with respect to
the distribution of impurities within a dynamic QD is
inevitable due to a SAW.
\medskip

In this paper, we consider a 1D Dirac-like electron gas in
a graphene nanoribbon in the presence of a SAW. We will introduce
two mechanisms for miniband formation. First, the nanoribbon
is modulated by a longitudinal potential from a SAW. Secondly,
the nanoribbon is periodically corrugated. We notice that the
second mechanism does not lead to a quantized current but
instead produces traps for Dirac electrons, thereby limiting
electron mobility. Our numerical calculations reveal
that localized electron trap states are an effective mechanism
to adversely affect the adiabatic transport because
the localized-trap levels lying within the minigaps are very
sensitive to the phase of either the SAW or the corrugation-induced
potential. Varying the weight or the position of the
$\delta$-potential leads to different positions of localized
trap levels within the minigaps of the nanoribbon.
Therefore, these inevitable fluctuations of the trap potential
in a realistic system would most likely impede the current
quantization.
\medskip

The rest of the paper is organized as follows. In Sec.\,\ref{sec2saw},
we present the formalism for calculating band structure with localized
trap states for nanoribbons in the presence of a SAW.
In Sec.\,\ref{sec3saw}, numerical results for nanoribbons in the
absence/presence of a SAW and those for corrugated nanoribbons
in the absence of a SAW are presented to demonstrate and explain
the localized trap states within the minigaps.
The conclusions drawn from these results are briefly summarized in Sec.\,\ref{sec4saw}.

\section{Miniband Structure with Localized Trap States}
\label{sec2saw}

The work done by Talyanskii, {\em et al.\/}\,\cite{11saw} on quantum
adiabatic charge transport focused on the coupling between
a semimetallic carbon nanotube and a SAW. The electron
backscattering from the SAW potential is used to induce a
miniband spectrum. The electron interactions enhance the minigaps
thereby improving current quantization. The effect due
to impurities in the carbon nanotube
is averaged by a  SAW potential.
\medskip

For the cases of a semimetallic carbon nanotube, semiconducting carbon
nanoribbon with applied SAW potential and corrugated nanoribbon, the energy
levels are given by the spectra of discretized 1D Dirac Hamiltonian
(see the Appendix\ A for detailed derivations)

\begin{equation}
\label{EQ:Ham}
\hat{\cal H} = \left[\begin{array}{*{20}c}
\ddots & \ddots & \ddots & \ldots &\ldots &\ldots &\ldots\\
\ddots & {a_{n - 1} } & {b_{n - 1}^ *  } & {c_{n - 1}^ *  } & 0 & 0 & 0  & \ldots\\
\ddots & {b_{n - 1} } & {a_{n - 1}^ *  } & 0 & {c_{n - 1} } & 0 & 0 & \ldots  \\
\ldots & {c_n } & 0 & {a_n } & {b_n^ *  } & {c_n^ *  } & 0 & \ldots \\
\ldots & 0 & {c_n^ *  } & {b_n } & {a_n^ *  } & 0 & {c_n } & \ldots \\
\ldots & 0 & 0 & {c_{n + 1} } & 0 & {a_{n + 1} } & {b_{n + 1}^ *  }  & \ddots\\
\ldots&  0 & 0 & 0 & {c_{n + 1}^ *  } & {b_{n + 1} } & {a_{n + 1}^ *  } & \ddots \\
\ldots& \ldots& \ldots & \ldots& \ldots & \ddots & \ddots & \ddots\\
\end{array}\right]_{2N \times 2N}\ .
\end{equation}
The eigenvalue problem is defined within the spatial interval
$0<x<2 \pi/k$ and assumes periodic boundary conditions.
In this notation, $k$ stands for either the wave number $k_{SAW}$
of the SAW potential or the wave number $k_c$ of an effective
potential induced by the corrugation. The discretization of the
Hamiltonian is provided by the mesh $x_n = n \delta x$ with
$n \in 0 \ldots N-1$ and $\delta x = 2 \pi / k N$.
\medskip

For either the carbon nanotube or graphene nanoribbon, the parameters
for the Hamiltonian matrix in Eq.\,\eqref{EQ:Ham} are given by

\begin{gather}
\label{EQ:CNano}
a_n = 0\ ,\\
\notag
b_n =  \Delta\,\texttt{e}^{-i\alpha(x_n)}\ ,\\
\notag
c_n = \frac{i \hbar\tilde{v}_F}{2\delta x}\ ,
\end{gather}
where we have introduced the SAW and impurities combined phase

\begin{equation}
\label{EQ:ALPHA}
\alpha(x) = \int\limits_{0}^{x} du\,\left[V_{SAW}(u)-V_{trap}(u)\right]
= \lambda \cos(k_{SAW}x) + V_0\,\theta(x-x_0)\ ,
\end{equation}
with $\lambda = 2A/(\hbar \tilde{v}_F k_{SAW})$  the normalized
SAW amplitude, $\tilde{v}_F$ is the Fermi velocity of graphene.
Additionally, $V_0=V_{trap}/(\hbar \tilde{v}_F k_{SAW})$ and $x_0$
denote the normalized trap-potential amplitude and position of the
short-range dynamic trap for electrons, respectively, and the trap
is sliding together with the SAW potential. The mass term involving
$\Delta$ is the original energy gap for the system in the absence of
a SAW. In case of a nanotube,  $\Delta$ may be generated by a
magnetic  field. For a nanoribbon, the gap is structural for the
semiconducting nanoribbon $\hbar \tilde{v}_F k_y^{(m)}$
with $k_y^{(m)}$ being the transverse electron wave number due to
finite size across the ribbon. For nanoribbons, the explicit form
for the phase introduced in Eq.\,\eqref{EQ:ALPHA} can be found from
the Appendix\ A.
\medskip

As far as the minigaps are concerned, the effect due to the SAW potential
on the nanoribbon may be compared with corrugation. A sinusoidal corrugated
semiconducting ribbon can be mapped on to a flat ribbon. The mapping introduces
an additional $\hat{\sigma}_1$ term into the Dirac equation, as described
in Appendix\ A. This yields

\begin{gather}
\label{EQ:Corug}
a_n=\hbar\tilde{v}_F\left[\frac{-\sqrt{2}{\cal C}^2 k_{c}^3\,\sin(2k_cx_n)}
{\Delta_c(x)[2+{\cal C}^2 k_c^2\,\cos(2k_cx_n)+{\cal C}^2k_c^2]^{3/2}}+i\frac{{\cal C}^2k_c^3}
{4\Delta_c^3(x)}\,\sin(2k_cx_n)\right]\ ,\\
\notag
b_n=i\hbar\tilde{v}_Fk_y^{(m)}\,\texttt{e}^{-2i\alpha_c(x_n)}\ ,\\
\notag
c_n=\frac{i\hbar\tilde{v}_F}{\Delta_c(x_n)\,2\delta x}\ ,
\end{gather}
where the corrugation-induced gap is given by
$\Delta_c(x_n)=\sqrt{1+{\cal C}^2k_c^2\cos^2(k_cx_n)}$,
with ${\cal C}$ being the normalized amplitude of the corrugation,
and $k_y^{(m)}$ is the quantized wave number across the nanoribbon.

\section{Numerical Simulation and Discussions}
\label{sec3saw}

\begin{figure}[ht]
\centering
\includegraphics[width=0.8\textwidth]{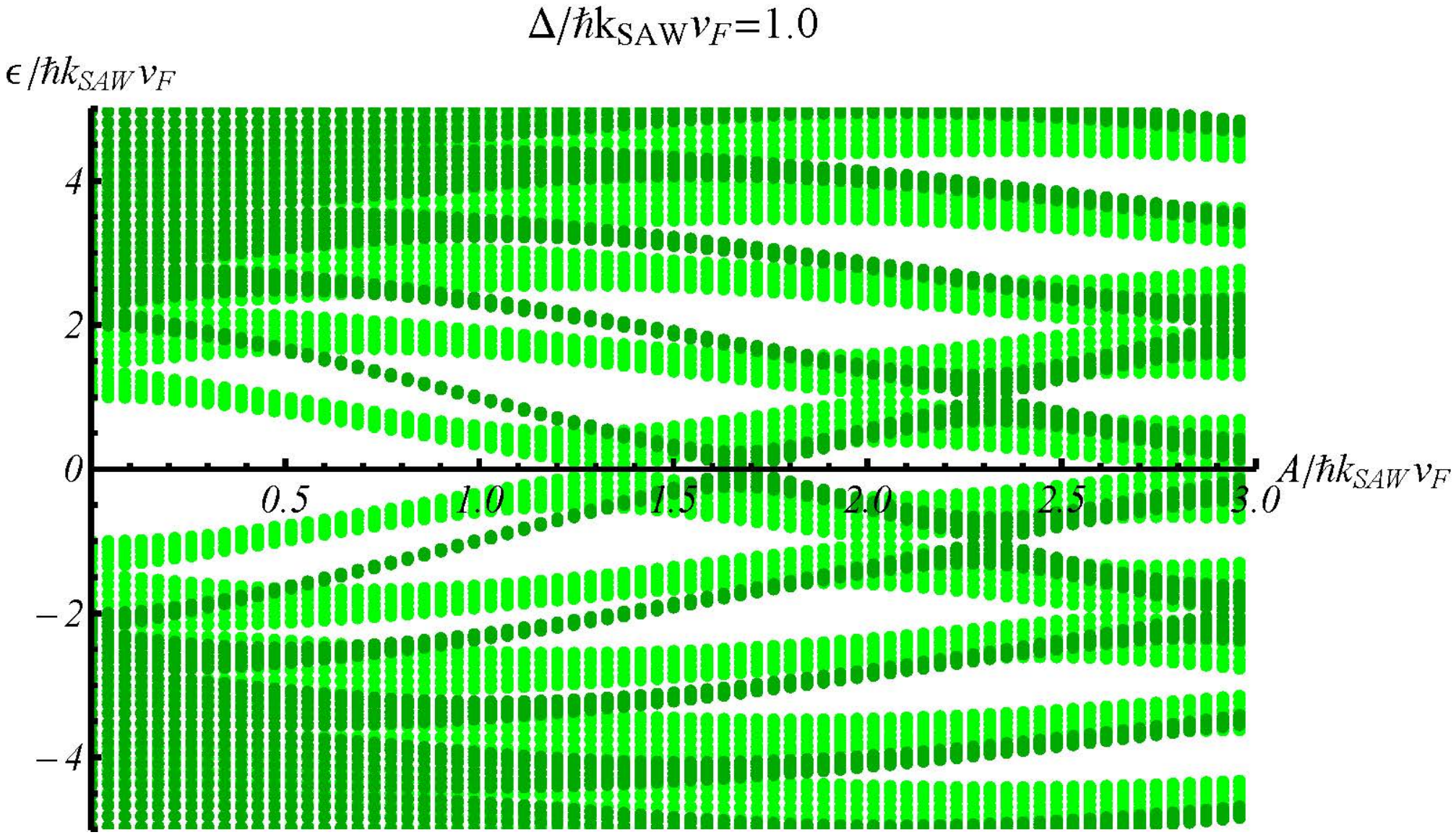}
\includegraphics[width=0.8\textwidth]{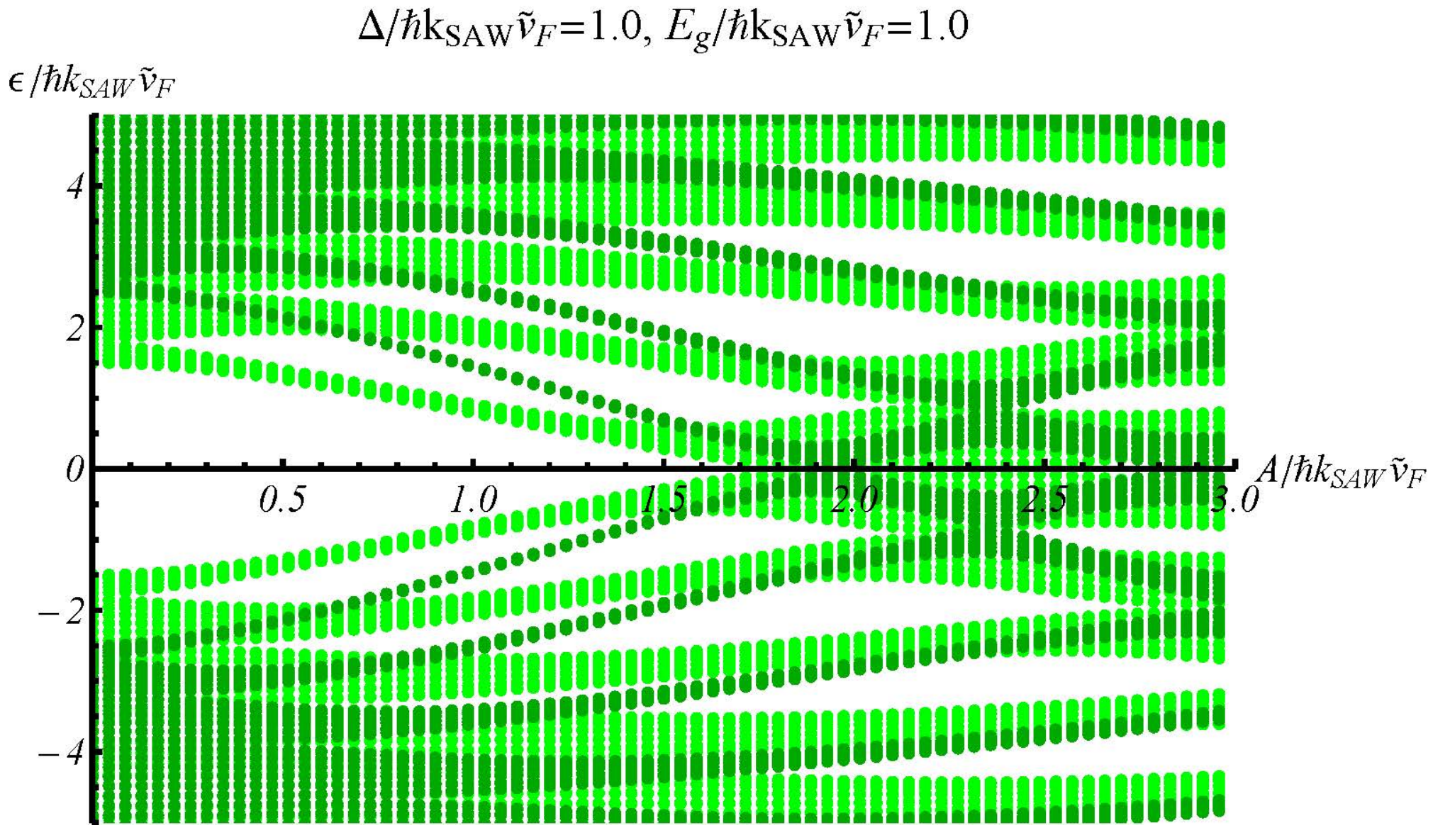}
\caption{\label{FIG:1} (Color online) Scaled electron
energy spectrum $\varepsilon(k)/(\hbar \tilde{v}_Fk_{SAW})$
in the absence of electron traps as a function of the
normalized SAW amplitude $A/(\hbar \tilde{v}_Fk_{SAW})$
for semiconducting nanoribbons subjected to an
acoustically induced SAW potential. In this figure,
we choose $\Delta/(\hbar \tilde{v}_Fk_{SAW})=1.0$
(upper panel) and $1.5$ (lower panel). Only the eigen-spectra
arising from the two lowest dispersion curves in the absence
of a SAW are displayed. Higher subbands contribute significantly
at larger SAW amplitude. The lighter shaded regions arise
from the lowest energy dispersion curves, whereas the
darker shaded regions are associated with the energy dispersions
of the next subband.}
\end{figure}

\begin{figure}[ht]
\centering
\includegraphics[width=0.4\textwidth]{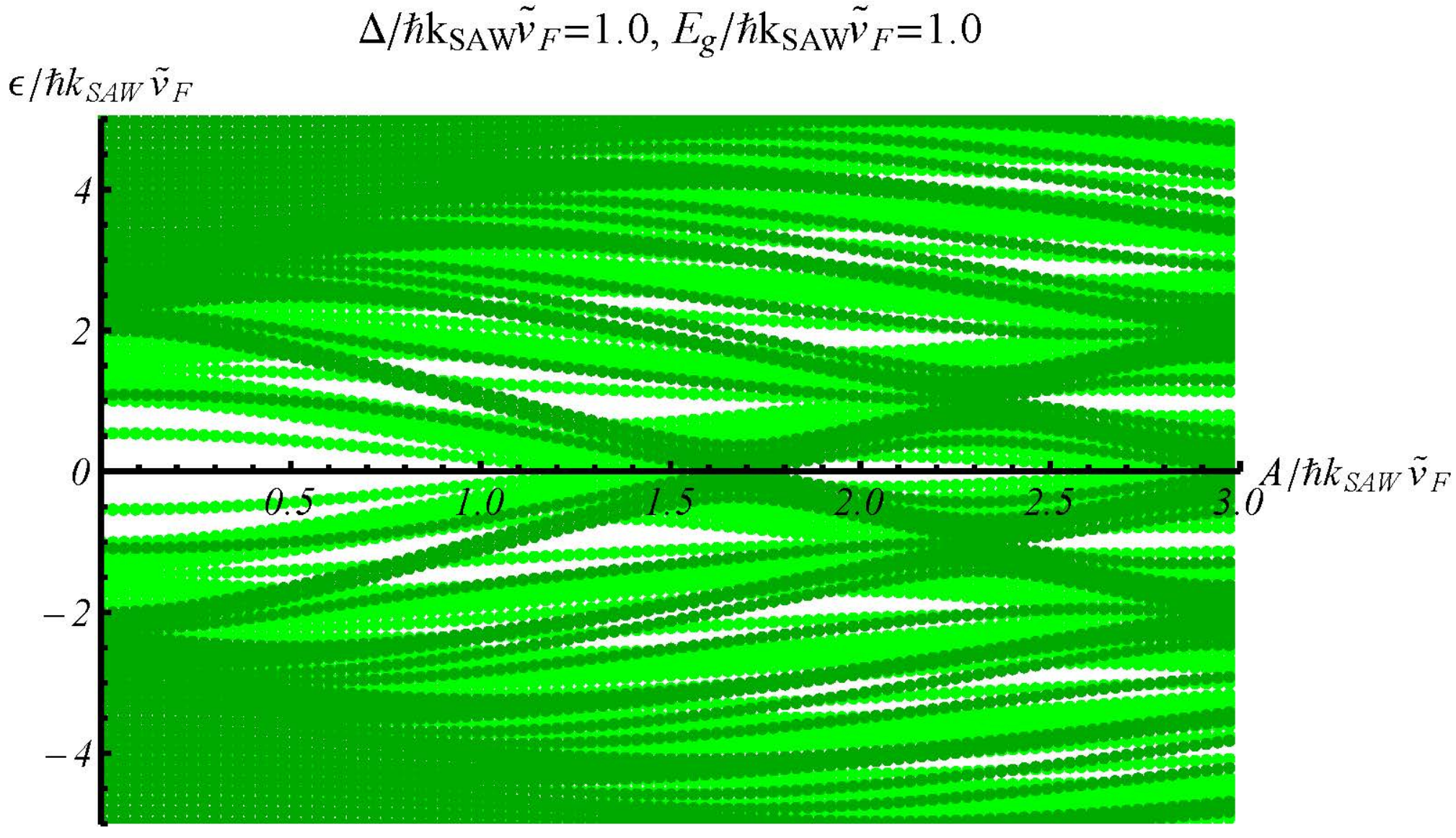}
\includegraphics[width=0.4\textwidth]{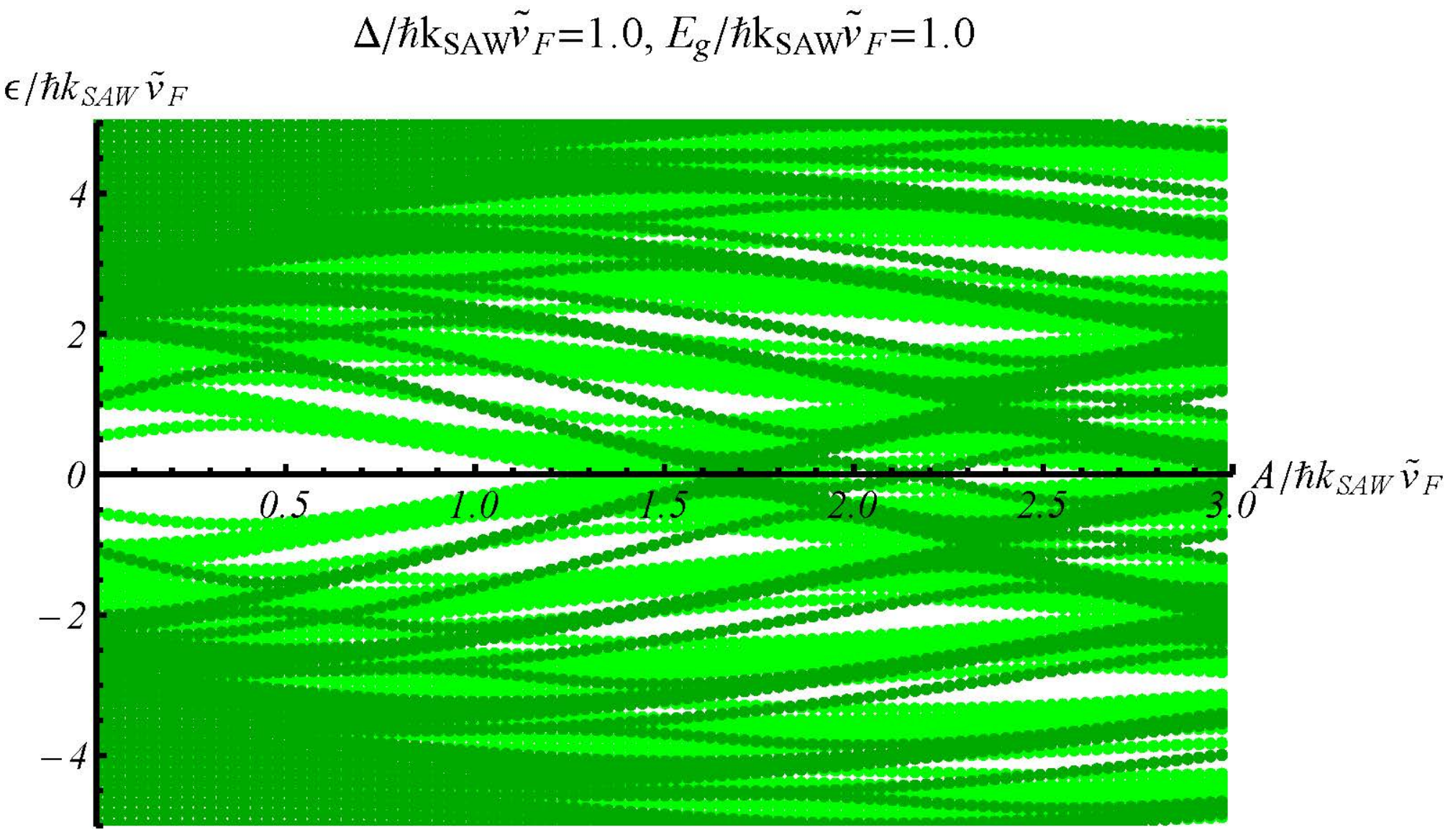}
\includegraphics[width=0.4\textwidth]{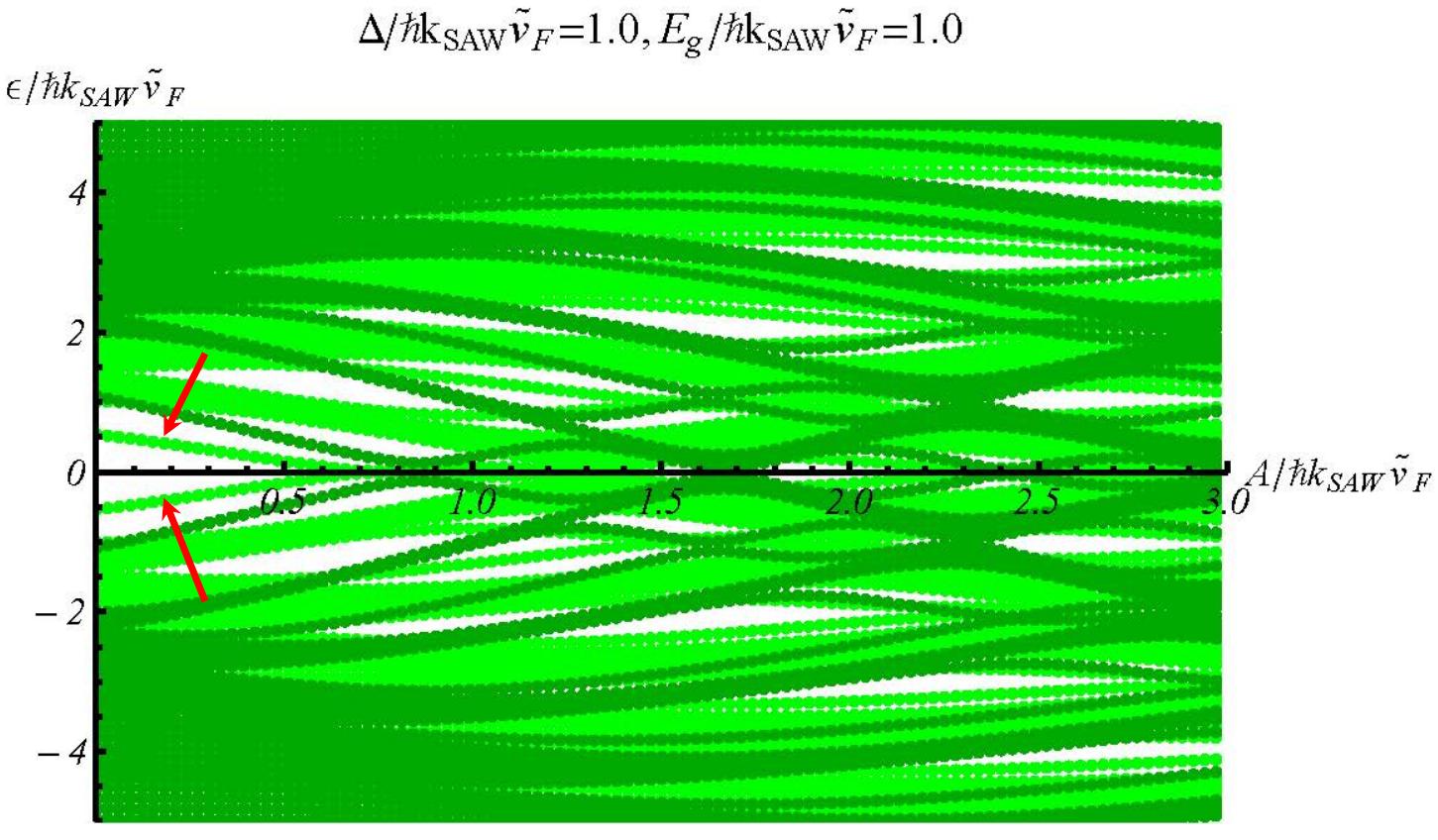}
\includegraphics[width=0.4\textwidth]{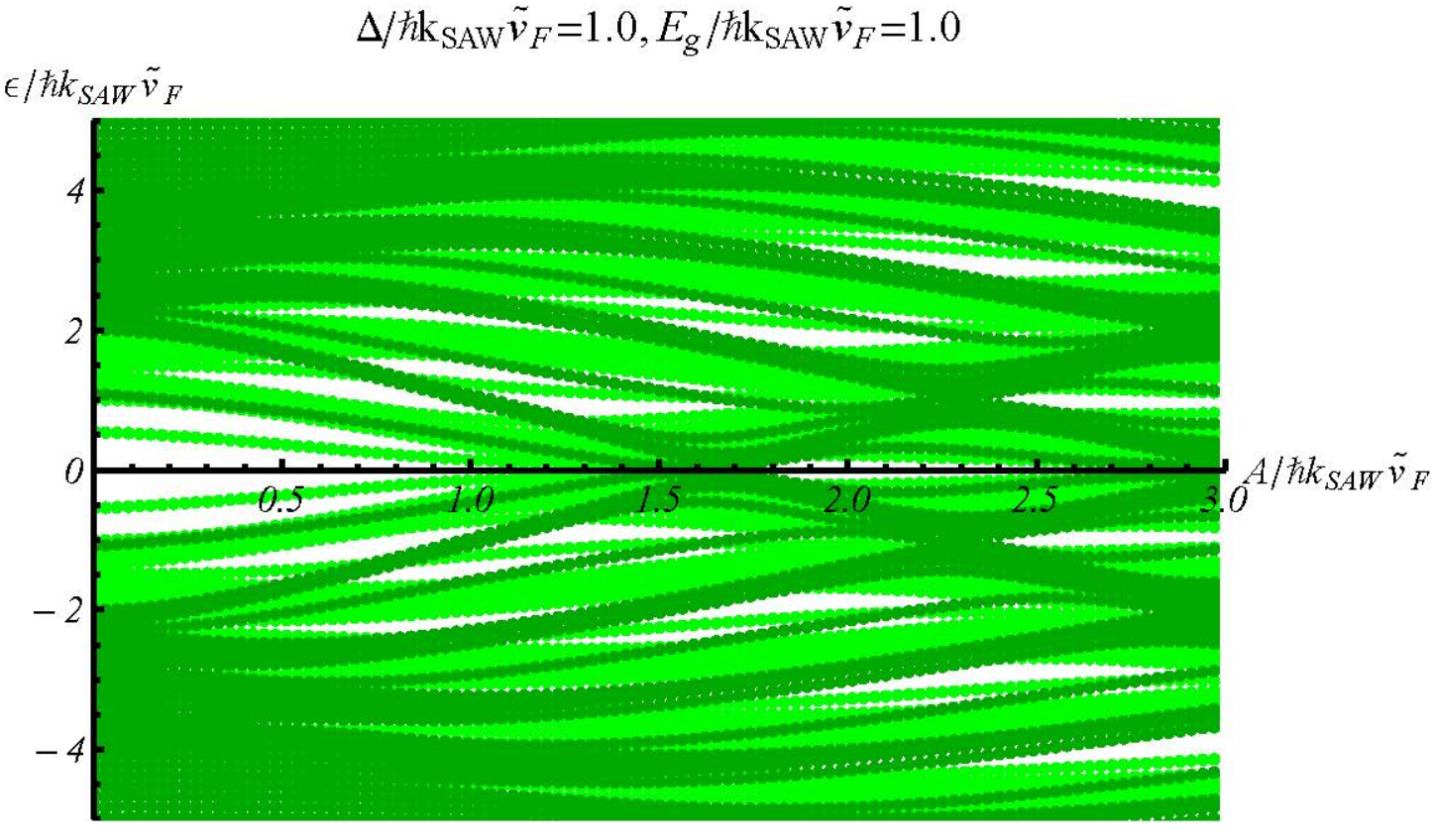}
\caption{\label{FIG:2} (Color online) Scaled electron
energy spectrum $\varepsilon(k)/(\hbar \tilde{v}_Fk_{SAW})$
in the presence of electron traps as a function of the
normalized SAW amplitude $A/(\hbar \tilde{v}_Fk_{SAW})$
for semiconducting nanoribbons subjected to an
acoustically induced SAW potential. In this figure, we chose
the energy gap $\Delta/(\hbar \tilde{v}_Fk_{SAW})=1.0$ and
the trap weight $V_0=V_{trap}/(\hbar \tilde{v}_Fk_{SAW})=1.0$
for all  four panels. As in Fig.\,\ref{FIG:1}, only the
eigen-spectra due to  the two lowest dispersion curves in
the absence of the SAW are shown. Here, the position
$x_0k_{SAW}/(2\pi)$ of  electron traps, moving with the SAW
potential, is located at: $1/3$ (upper-left panel), $1/5$
(lower-left panel), $1/7$ (upper-right panel), and $1/11$
(lower-right panel). Two induced trap-state energy levels
within the energy gap are indicated by arrows in the lower-left
panel for emphasis.}
\end{figure}

\begin{figure}[ht]
\centering
\includegraphics[width=0.8\textwidth]{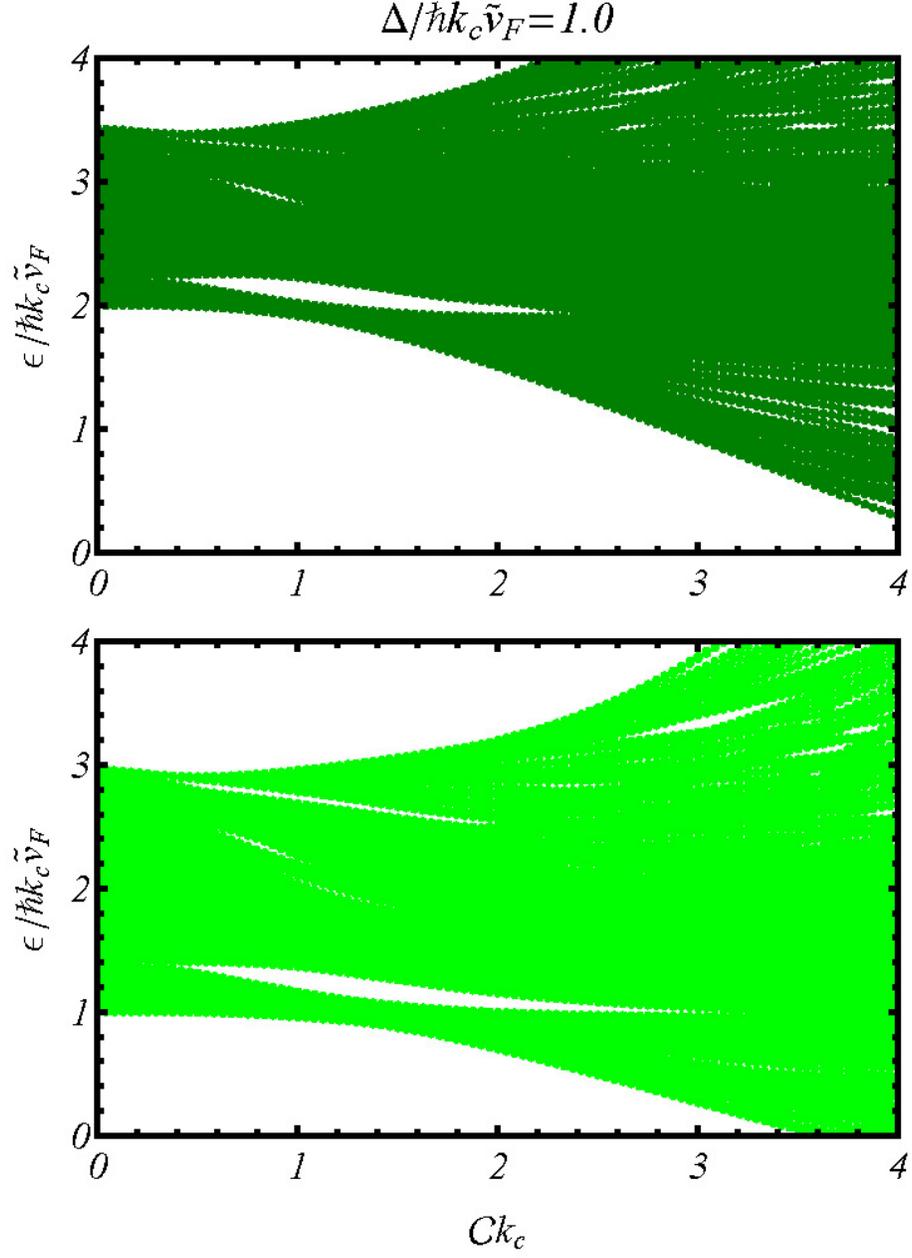}
\caption{\label{FIG:3} (Color online) Minigap spectrum
$\varepsilon(k)/(\hbar \tilde{v}_Fk_c)$, induced by a
corrugation potential, as a function of the normalized
modulation amplitude ${\cal C}k_c$. In this figure, the two
plots correspond to the two lowest values of the quantized
transverse wave number $k_y^{(m)}$.  The graph at the top
comes from the lowest quantized energy subband, whereas the graph
at the  bottom is due to   the first-excited subband.}
\end{figure}

In our numerical calculations, all the energies in Figs.\,\ref{FIG:1}
and \ref{FIG:2}, such as $\varepsilon$, $\Delta$ and $E_g$,
are normalized to $\hbar \tilde{v}_F k_{SAW}$.
The SAW potential amplitude $A$ is also normalized to
$\hbar \tilde{v}_F k_{SAW}$. Additionally, all the energies
in Fig.\,\ref{FIG:3}, such as $\varepsilon$ and $\Delta$, are
measured in units of   $\hbar\tilde{v}_F k_{c}$, and the
corrugation  amplitude ${\cal C}$ is normalized to $1/ k_{c}$.
Besides, the transverse wave number $k^{(m)}_y$ in all the
plots is scaled by $2\pi/3a_0$. In this way, we are able to
draw some universal conclusions concerning the effects
due to  minigaps.
\medskip

In Fig.\,\ref{FIG:1}, we compare the energy band structure
of nanoribbons for two values of $\Delta$ in the absence of
electron traps. Two values of $k_y^{(m)}$ were chosen
(light and dark colors) to describe the two lowest energy
levels (see the discussions in the Appendix\ A). The minigaps are generated by a sliding dynamic QD and they oscillate as a function of the SAW
amplitude $A$, as may be verified using perturbation theory,
vanishing at values close, but generally not equal, to the
roots of Bessel functions. Increasing or decreasing the value
of $\Delta$ results in a shift of the nodes on the graph as
evidenced by comparing our results in Fig.\,\ref{FIG:1}.
Therefore, $\Delta$ determines not only the magnitude of
the original gap in the absence of a SAW but also the size
of the minigaps in the presence of a SAW. Higher energy minigaps
are partially closed by the energy levels corresponding
to a larger value of $k_y^{(m)}$ (not shown here).
\medskip

We now introduce electron traps into our nanoribbon by superposing
a negative $\delta$-potential onto the  SAW potential so as
to simulate a short-range Coulomb interaction. In this case,
the position of the trap is fixed in the moving SAW frame of
reference, which is quite different from embedded impurities in
a nanostructure. In the moving SAW frame, the embedded impurities
are moving against the dynamic QDs created by both the transverse
dimension of the nanostructure  and the longitudinal SAW potential.
This results in an average of the impurity effects
with respect to these dynamic QDs in the longitudinal direction.
As seen in the results presented in Fig.\,\ref{FIG:2}, localized
trap states occur within the minigaps once the weight of
the trap potential $V_0$ becomes larger than
$\Delta/(\hbar \tilde{v}_Fk_{SAW})$. Relative energy value of these
trap states in the presence of SAW is sensitive to the position $x_0$ of the
trap within a dynamic QD. If we set $\lambda=V_0$, then the
contribution to $\alpha(x)$ from the trap located in the nodes
of the SAW potential is fully compensated by the cosine term in Eq.\,\eqref{EQ:ALPHA}. As a result of this compensation, the
localized trap states will disappear from the  gap and minigap
regions. If we extend the single-trap model employed in
this paper to a uniform distribution of traps, the fluctuations
in the phase term $\alpha(x)$ [see Eq.\,\eqref{EQ:ALPHA}] would fill up the entire minigap region with
a delocalized trap band. Consequently, the adiabatic
approximation may not be applicable. In other words, to satisfy
the adiabatic assumption, one must have dominance of the SAW
potential, i.e., $\lambda \gg V_0$ must be satisfied.
\medskip

We compare the results for SAW-based dynamic QDs
in Figs.\,\ref{FIG:1} and \ref{FIG:2} with those for  static
QDs created by  corrugation on a graphene nanoribbon in the
absence of a SAW and electron traps. This we do by
displaying in Fig.\,\ref{FIG:3} the minigaps induced by
the  corrugation. We find from the figure that minigaps
only exist for finite but small values of the corrugation amplitude
${\cal C}$. This means that  these minigaps are generally
much less  than those induced by a SAW.
As a matter of  fact, the existence of non-vanishing
diagonal terms $a_n$ given  in Eq.\,\eqref{EQ:Corug},
effectively mitigates the phase fluctuations in the
off-diagonal terms $b_n$. This keeps the minigaps open
and the energy spectra robust even after traps have been introduced to cause a fluctuation in the phase term $\alpha_c(x)$.
\medskip

Finally, let us assume that a narrow channel is formed
within  a two-dimensional electron-gas layer lying in
the $xy$-plane. We will neglect the finite thickness of the
quantum well for the heterostructure in the $z$-direction
and consider the electron motion as strictly two dimensional.
We will employ one of the simplest models for the gate-induced or etched\,\cite{16saw} confining electrostatic potential.
In this way, a 1D channel is formed on the two-dimensional
electron-gas layer and the  dynamics of massive electrons
can be modeled by a discretized 1D Schrodinger equation.
However, from numerical  results (not shown here), we find
no evidence of the
minigaps for this model, i.e., the minigaps are the
characteristics of Dirac fermions.

\section{Concluding Remarks}
\label{sec4saw}

In conclusion, we have calculated in this paper the energy
band structure for graphene nanoribbons, embedded with
a single electron trap, upon which a SAW is launched.
Our results show that localized trap states appear in the
minigaps. More importantly, the location of the trap
state-energy level is determined by the positions of
the trap with respect to the phase of the sinusoidal SAW.
Consequently, the adiabatic approximation might not be
appropriate whenever the minigap is less or comparable
with the weight of a short-range $\delta$-potential for
the trap (see Fig.\,\ref{FIG:2}). On the other hand, in
Fig.\,\ref{FIG:1}, where there are no electron traps,
a larger value of $\Delta$ in the energy spectrum leads to
a substantial increase in the number of minigaps as well as
the ballistic current quantization. Periodic corrugation
of the nanoribbon may be used instead of a SAW as a mechanism
for inducing minigaps. Those are expected to be less sensitive
to the presence of charged impurities or electron trap potentials.

\acknowledgments
The authors are very grateful to Professor Leonid Levitov
for helpful suggestions and critical comments during the
course of this work. His critical remarks have undoubtedly
helped to strengthen the presentation of this paper. This research
was supported by the contract \# FA 9453-07-C-0207 of AFRL. DH would like
to thank the Air Force Office of Scientific Research (AFOSR)
for its support.

\newpage
\begin{appendix}

\section{Energy Band Calculations in Absence of Electron Traps}
\vspace{0.2cm}

\subsection{Carbon Nanotubes}

The electron eigenstates in a semi-metallic nanotube are
described by a 1D Dirac equation. For simplicity, a
noninteracting system is considered here. Under the
stationary approximation, the single particle energy spectrum
$\varepsilon(k)$ is obtained from the following perturbed 1D Dirac
equation

\begin{gather}
\label{EQ:1_1}
\varepsilon(k)\,\psi_\alpha(x) = - i \hbar \tilde{v}_F\,\frac{\partial\psi_\alpha(x)}
{\partial x}  + \Delta\,\psi_\beta(x) + A \sin (kx)\,\psi_\alpha(x)\ ,\\
\label{EQ:1_2}
\varepsilon(k)\,\psi_\beta(x) =  i \hbar  \tilde{v}_F\,
\frac{\partial\psi_\beta(x)}{\partial x}  +
\Delta\,\psi_\alpha(x) +   A \sin (kx)\,\psi_\beta(x)\ .
\end{gather}
In this notation, $k$ represents the electron wave number
along the nanotube, $\tilde{v}_F$ is the Fermi velocity of
Dirac electrons, $A$ is the SAW amplitude, $\Delta$ is the
energy gap of the system in the absence of a SAW, $\alpha$ and
$\beta$ label the two sublattices of graphene from which
the nanotube is rolled. In addition, we require $k=k_{SAW}$
to satisfy momentum conservation, where $k_{SAW}$ is the
wave number of a SAW propagating along the nanotube. To explore
the miniband structure due to quantum confinement in the radial
direction, a gauge transformation is implemented and is defined by

\begin{gather}
\label{EQ:1_3}
\left[{
\begin{array}{c}
\psi_\alpha(x)\\
\psi_\beta(x)
\end{array}
}\right]
=
\texttt{e}^{(i/2) \hat{\sigma}_3 \lambda \cos(k x)}
\left[{
\begin{array}{c}
\psi^\prime_\alpha(x)\\
\psi^\prime_\beta(x)
\end{array}
}\right] =
\left[{
\begin{array}{c}
\texttt{e}^{(i/2) \lambda \cos(kx)}\,\psi^\prime_\beta(x)\\
\texttt{e}^{(-i/2) \lambda \cos(kx)}\,\psi^\prime_\alpha(x)
\end{array}
}\right]  \ ,
\end{gather}
where $\lambda = 2A/(\hbar \tilde{v}_F k)$ is the normalized SAW
amplitude. Substituting Eq.\,\eqref{EQ:1_3}
into Eqs.\,\eqref{EQ:1_1} and \eqref{EQ:1_2}, we obtain

\begin{gather}
\varepsilon(k)\,\texttt{e}^{(i/2) \lambda \cos(kx)}\,\psi^\prime_\beta(x) = - i \hbar \tilde{v}_F \frac{\partial}{\partial x} \left[{\texttt{e}^{(i/2) \lambda \cos(kx)}\,\psi^\prime_\beta(x)}\right]\\
\notag
+\Delta\,\texttt{e}^{(-i/2) \lambda \cos(kx)}\,\psi^\prime_\alpha(x) + A \sin (kx)\,\texttt{e}^{(i/2) \lambda \cos(kx)}\,\psi^\prime_\beta(x)\ ,\\
\label{EQ:1_4}
\varepsilon(k)\,\texttt{e}^{(-i/2) \lambda \cos(kx)}\,\psi^\prime_\alpha(x)= i \hbar  \tilde{v}_F \frac{\partial}{\partial x} \left[{\texttt{e}^{(-i/2) \lambda \cos(kx)}\,\psi^\prime_\alpha(x)}\right]\\
\notag
+\Delta\,\texttt{e}^{(i/2) \lambda \cos(kx)}\,\psi^\prime_\beta(x) +   A \sin (kx)\,\texttt{e}^{(-i/2) \lambda \cos(kx)}\,\psi^\prime_\alpha(x)\ .
\label{EQ:1_5}
\end{gather}
By introducing the following identities

\begin{gather}
\label{EQ:1_6}
- i \hbar \tilde{v}_F\,\frac{\partial}{\partial x}
\left[{\texttt{e}^{(i/2) \lambda \cos(kx)}\,\psi^\prime_
\beta(x)}\right] = \\
\notag
-A \sin(k x)\,\psi^\prime_\beta(x) - i \hbar \tilde{v}_F\,
\texttt{e}^{(i/2) \lambda \cos(k x)}\,
\frac{\partial\psi^\prime_\beta(x) }{\partial x}\ ,\\
\label{EQ:1_7}
 i \hbar \tilde{v}_F\,\frac{\partial}{\partial x}
 \left[{\texttt{e}^{(-i/2) \lambda
\cos(kx)}\,\psi^\prime_\alpha(x)}\right] = \\
\notag
-A \sin(k x)\,\psi^\prime_\alpha(x) + i \hbar \tilde{v}_F\,
\texttt{e}^{(-i/2) \lambda \cos(k x)}\,
\frac{\partial\psi^\prime_\alpha(x)}{\partial x}\ ,
\end{gather}
Eqs.\,\eqref{EQ:1_4} and \eqref{EQ:1_5} may be simplified as

\begin{gather}
\label{EQ:1_8}
\varepsilon(k)\,\psi^\prime_\beta(x)=
- i \hbar \tilde{v}_F\,\frac{\partial\psi^\prime_\beta(x)}{\partial x}
+ \Delta\,\texttt{e}^{-i \lambda \cos(kx)}\,\psi^\prime_\alpha(x)\ ,\\
\label{EQ:1_9}
\varepsilon(k)\,\psi^\prime_\alpha(x)=
i \hbar  \tilde{v}_F\,\frac{\partial\psi^\prime_\alpha(x)}{\partial x}
+\Delta\,\texttt{e}^{i \lambda \cos(kx)}\,\psi^\prime_\beta(x)\ .
\end{gather}
Furthermore, by employing the basis set $\hat{\Psi}^\prime(x)\equiv
\left\{{\psi^\prime_\alpha(x),\,\psi^\prime_\beta(x)}\right\}$,
the above equations can be rewritten into a compact matrix form, given by

\begin{gather}
\label{EQ:1_10}
\varepsilon(k)\,\hat{\mathcal{I}}\,\hat{\Psi}^\prime(x) =
\hat{\mathcal{H}}\,\hat{\Psi}^\prime(x)\ ,\\
\label{EQ:1_11}
\hat{\mathcal{H}} =
 i \hbar \tilde{v}_F \hat{\sigma_3}\,\frac{\partial}{\partial x} +
\Delta\,\hat{\sigma}_1\,\texttt{e}^{-i \lambda \hat{\sigma}_3 \cos(k x)} \ .
\end{gather}

We will solve the eigenvalue problem within the spatial interval
$0<x<2 \pi/k$ and introduce the $N$-point mesh $x_n = n\,\delta x$, where
$n \in 0 \ldots N-1$ and $\delta x = 2 \pi / k N$. In this way,
the derivative on the mesh can be approximated by
$\partial\hat{\Psi}^\prime(x_n)/\partial x\to
[\hat{\Psi}^\prime(x_{n+1})-\hat{\Psi}^\prime(x_{n-1})]/(2
\delta x)$. Especially, on this spatial mesh,
the Hamiltonian in Eq.\,\eqref{EQ:1_11} may be projected into
the matrix given in Eq.\,\eqref{EQ:Ham}.

\subsection{Graphene Nanoribbons}

Here, we consider an armchair graphene nanoribbon lying
along the $x$-direction. The total number of carbon atoms
(in both sublattices) across the ribbon is assumed to be
$M$. The armchair edges mix up the graphene $K$ and $K^\prime$
valleys so that the wave function becomes

\begin{equation}
\label{EQ:2_13}
\hat{\Psi}_{m}(x,\,y) = \texttt{e}^{i k_y^{(m)} y}\,
\hat{\Psi}_{K,\,m}(x)  +  \texttt{e}^{-i k_y^{(m)} y}\,
\hat{\Psi}_{K^\prime,\,m}(x)\ ,
\end{equation}
where the transverse wave number is given by

\begin{equation}
\label{EQ:2_14}
k_y^{(m)} = \frac{2 \pi m}{2L + a_0} + \frac{2 \pi}{3 a_0}\ ,
\end{equation}
$m=0,\,\pm 1,\,\pm 2,\,\cdots$ is an integer, $L$ is the nanoribbon width, and $a_0=\sqrt{3}a/2$ ($a\approx 1.42$\AA) is the size of the unit cell in graphene.
Nanoribbons with width $L/a_0= 3M +1$ give rise to the following relation

\begin{gather}
\label{EQ:2_15}
k_y^{(m)} = \frac{2 \pi m}{6 M a_0 + 3 a_0}+ \frac{2 \pi}{3 a_0}
= \frac{2 \pi}{3 a_0}\left({\frac{2 M + 1 + m}{2 M+1}}\right)\ ,
\end{gather}
and it is clear that the minimum energy occurs at $k_y^{(-2M-1)} = 0$.
Therefore, such nanoriibons are metallic. The next miniband corresponds
to $k_y^{(-2M)} =(2 \pi/3 a_0)\,[1/(2 M+1)]$. On the other hand,
for nanoriibons having width $L/a_0 = 3M$ (upper Eq.) or $L/a_0=3M-1$
(lower Eq.), the two minimal values of $k_y^{(m)}$ are found to be

\begin{gather}
\label{EQ:2_16}
\begin{array}{cc}
k_y^{(-2M)} = (2 \pi/3 a_0)\,[1/(6M +1)]\ \ \ \mbox{and}\ \ & k_y^{(-2M-1)}
=(2 \pi/3 a_0)\,[-2/(6M +1)]\ ,\\
k_y^{(-2M)} = (2 \pi/3 a_0)\,[-1/(6M -1)]\ \ \ \mbox{and}\ \ & k_y^{(-2M+1)}
=(2 \pi/3 a_0)\,[2/(6M -1)]\ .
\end{array}
\end{gather}
Those nanoribbons are semiconducting with the energy gap determined by

\begin{equation}
\label{EQ:2_17}
\Delta(M) = \frac{2 \pi \hbar \tilde{v}_F }{3 a_0}
\left({\frac{1}{6M \pm 1}}\right) \ .
\end{equation}
The $x$-component of the wave function in Eq.\,\eqref{EQ:2_13}
may be determined by

\begin{gather}
\label{EQ:2_18}
\varepsilon(k)\,\hat{\mathcal{I}}\hat{\Psi}_{K(K^\prime),\,m}(x) =
\hat{\mathcal{H}}\,\hat{\Psi}_{K(K^\prime),\,m}(x)\ ,\\
\label{EQ:2_19}
\hat{\mathcal{H}} = \pm
\hbar \tilde{v}_F \left(-i \hat{\sigma}_1 \frac{\partial}{\partial x} + i k_y^{(m)}\,\hat{\sigma}_2\right)+
A \sin(k x)\,\hat{\mathcal{I}} \ ,
\end{gather}
where the $\pm$ signs correspond to $K$ and $K^\prime$ valleys. Additionally,
the Hamiltonian in Eq.\,\eqref{EQ:2_19} can be transformed into the
form in Eq.\,\eqref{EQ:1_11} after applying the following
unitary transformation

\begin{equation}
\label{EQ:2_20}
\hat{\mathcal{U}} =
\left[{
\begin{array}{cc}
\texttt{e}^{-i \alpha(x)} & -\texttt{e}^{i \alpha(x)}\\
\texttt{e}^{-i \alpha(x)} & \texttt{e}^{i \alpha(x)}
\end{array}
}\right] \ ,
\end{equation}
where $\alpha(x) = -A \cos(k_{SAW}x)/(\hbar
\tilde{v}_F k_{SAW})$. As a result,
the transformed Hamiltonian takes the form

\begin{gather}
\label{EQ:2_21}
\hat{\cal H}^\prime = \hat{\cal U}^\dag\otimes\hat{\cal H}\otimes\hat{\cal U}\\
\notag
=\pm \hbar \tilde{v}_F
\left[\begin{array}{cc}
- i \partial/\partial x & -i k_y^{(m)} \texttt{e}^{i 2 \alpha(x)}\\
i k_y^{(m)} \texttt{e}^{-i 2 \alpha(x)} &  i \partial/\partial x
\end{array}\right]\ .
\end{gather}
Formally, this Hamiltonian is equivalent to that in
Eq.\,\eqref{EQ:1_11} after we applying the following substitutions

\begin{gather}
\label{EQ:2_22}
2\alpha(x)\to -\lambda \cos(k_{SAW}x)\\
\notag
\hbar \tilde{v}_F k_y^{(m)}\to \Delta\ .
\end{gather}
The valley sign $\pm$ does not change anything due to
the mirror symmetry in the energy dispersion relation
$\varepsilon(k)$ with respect to $\pm k$.

\subsection{Corrugated Nanoribbons}

We now turn to the case of  a corrugated graphene nanoribbon
whose modulation is sinusoidal with  amplitude
${\cal C}$ and wavelength $2\pi/k_c$ along the $x$-axis.
The model Hamiltonian for such a corrugated graphene
nanoribbon has been given in Ref.\,[\onlinecite{PRB-2010}] as

\begin{equation}
\hat{\cal H} =  - i\hbar\tilde{v}_F\hat{\sigma _1}\,\frac{\partial}
{\partial x} + \hbar\tilde{v}_Fk_y\hat{\sigma_2} - i\hbar\tilde{v}_F\hat{\sigma_1}\,\frac{K(x)}{2}\,\frac{df(x)/dx}{\Delta_c(x)}\ ,
\end{equation}
where

\[
\begin{gathered}
\Delta_c(x)   = \sqrt {1 + [df(x)/dx]^2 }\ \ \ \mbox{and}\ \ \ K(x) = \frac{-d^2f(x)/dx^2}
{1 + [df(x)/dx]^2}\ , \hfill \\
f(x)= {\cal C}\sin(k_cx)\ . \hfill \\
\end{gathered}
\]
After applying the following unitary transformation

\[
\hat{\cal U} = \frac{1}
{{\sqrt 2 }}\left[ {\begin{array}{*{20}c}
{\texttt{e}^{ - i\alpha_c(x)}} & { -\texttt{e}^{i\alpha_c(x) }}  \\
{\texttt{e}^{ - i\alpha_c(x)}} & {\texttt{e}^{i\alpha_c(x)}}
\end{array} } \right]\ \ \ \mbox{and}\ \ \ \hat{\cal U}^\dag   = \frac{1}
{{\sqrt 2 }}\left[ {\begin{array}{*{20}c}
{\texttt{e}^{i\alpha_c(x)}} & {\texttt{e}^{i\alpha_c(x)}}  \\
{ - \texttt{e}^{ - i\alpha_c(x)}} & {\texttt{e}^{ - i\alpha_c(x)}}  \\
 \end{array} } \right]\ ,
\]
where

\[
\begin{gathered}
\alpha_c(x)  = \frac{{{\cal C}^2 k_{c} ^3 }}
{4}\int\limits_0^x du\, {\frac{{\sin \left( {2k_{c} u} \right)}}
{{\left[ {1 + {\cal C}^2 k_{c}^2 \cos ^2 \left( {k_{c} u}
\right)} \right]^{3/2} }}}  \hfill \\
= \frac{1}
{2}\left[ {\frac{{\sqrt 2 }}
{{\sqrt {2 + {\cal C}^2 k_{c}^2 \cos \left( {2k_{c} x} \right)
+ {\cal C}^2 k_{c}^2 } }} - \frac{1}
{{\sqrt {1 + {\cal C}^2 k_{c}^2 } }}} \right]\ , \hfill
\end{gathered}
\]
the transformed Hamiltonian becomes

\begin{equation}
\hat{\cal H}^\prime = \hbar\tilde{v}_F\left[ {\begin{array}{*{20}c}
G(x) - \frac{i\partial/\partial x}
{{\Delta_c(x)}} & { - ik_y \texttt{e}^{2i\alpha_c(x)}}  \\
{ik_y \texttt{e}^{ - 2i\alpha_c } } & G^\ast(x) + \frac{i\partial/\partial x}
{{\Delta_c(x)}}
\end{array} }\right]\ ,
\label{add1}
\end{equation}
where the complex function is defined by

\[
G(x) = \frac{-\sqrt{2}{\cal C}^2k_c^3\sin(2k_cx)}
{\Delta_c(x)\left[2+{\cal C}^2k_c^2
\cos(2k_cx)+{\cal C}^2k_c^2\right]^{3/2}}
+ i\frac{{\cal C}^2k_c^3}
{4\Delta_c^3(x)}\,\sin(2k_cx)\ .
\]
By making use of the finite-difference method for calculating
$\partial/\partial x$ along with the following basis set

\begin{equation}
\varphi  = \left\{ {A_{n - 1} ,B_{n - 1} ,A_n ,B_n ,A_{n + 1} ,
B_{n + 1} } \right\}\ ,
\end{equation}
the transformed Hamiltonian matrix in Eq.\,\eqref{add1} may
be projected as

\[
\hat{\cal H} = \hbar\tilde{v}_F\left[\begin{array}{*{20}c}
G(x_{n-1}) & -ik_y\texttt{e}^{2i\alpha_c(x_{n-1})} & \frac{-i}{\Delta_c(x_{n-1})\,2\delta x} & 0 & 0 & 0\\
ik_y\texttt{e}^{-2i\alpha_c(x_{n-1})} & G^\ast(x_{n-1}) & 0 & \frac{i}{\Delta_c(x_{n-1})\,2\delta x} & 0 & 0\\
\frac{i}{\Delta_c(x_n)\,2\delta x} & 0 & G(x_n) &
-ik_y\texttt{e}^{2i\alpha_c(x_n)} & \frac{-i}{\Delta_c(x_n)\,2\delta x} & 0\\
0 & \frac{-i}{\Delta_c(x_n)\,2\delta x} & ik_y\texttt{e}^{-2i\alpha_c(x_n)} & G^\ast(x_n) & 0 & \frac{i}{\Delta_c(x_n)\,2\delta x}\\
0 & 0 & \frac{i}{\Delta_c(x_{n+1})\,2\delta x} & 0 & G(x_{n+1}) & -ik_y\texttt{e}^{2i\alpha_c(x_{n+1})}\\
0 & 0 & 0 & \frac{-i}{\Delta_c(x_{n+1})\,2\delta x} & ik_y\texttt{e}^{-2i\alpha_c(x_{n+1})} & G^\ast(x_{n+1})\\
\end{array}\right]\ .
\]
The above Hamiltonian matrix has the same form as Eq.\,\eqref{EQ:Ham}.
\end{appendix}

\newpage

\end{document}